\documentclass[twocolumn,aps,superscriptaddress]{revtex4}
\usepackage{epsfig}
\newcommand{\be}{\begin{equation}}
\newcommand{\ee}{\end{equation}}
\newcommand{\bea}{\begin{eqnarray}}
\newcommand{\eea}{\end{eqnarray}}

\newcommand{\bc}{\begin{center}}
\newcommand{\ec}{\end{center}}

\newcommand{\forget}[1]{}

\begin{document}

\preprint{}
\title{High-fidelity linear optical quantum computing with polarization encoding}
\author{Federico M. Spedalieri}
\email{Federico.Spedalieri@jpl.nasa.gov}
\affiliation{Jet Propulsion Laboratory,
California Institute of Technology, Mail Stop 126-347, 4800 Oak Grove Drive, 
Pasadena, California 91109-8099}
\author{Hwang Lee} 
\affiliation{Jet Propulsion Laboratory,
California Institute of Technology, Mail Stop 126-347, 4800 Oak Grove Drive, 
Pasadena, California 91109-8099}
\affiliation{Hearne Institute for Theoretical Physics, 
Department of Physics and Astronomy, Louisiana State University,
Baton Rouge, Louisiana 70803-4001}
\author{Jonathan P. Dowling}
\affiliation{Jet Propulsion Laboratory,
California Institute of Technology, Mail Stop 126-347, 4800 Oak Grove Drive, 
Pasadena, California 91109-8099}
\affiliation{Hearne Institute for Theoretical Physics, 
Department of Physics and Astronomy, Louisiana State University,
Baton Rouge, Louisiana 70803-4001}
\affiliation{Institute for Quantum Studies, Department of Physics, Texas A\&M
University, College Station, Texas 77843-4242}
\date{\today}

\begin{abstract}
We show that the KLM scheme [Knill, Laflamme and Milburn, Nature {\bf 409}, 46] can
be implemented using polarization encoding, thus reducing the number of path modes required
by half. One of the main advantages of this new implementation is that 
it naturally incorporates a loss detection mechanism that makes the
probability of a gate introducing a non-detected error, when non-ideal detectors are considered,
dependent only on the detector dark-count rate and independent of its efficiency.
Since very low dark-count rate detectors are currently available, a high-fidelity gate (probability
of error of order $10^{-6}$ conditional on the gate being successful) can be implemented
using polarization encoding. The detector efficiency determines the overall success probability
of the gate but does not affect its fidelity. This can be applied to the efficient 
construction of optical cluster states with very high fidelity for quantum computing.
\end{abstract}

\maketitle

\section{Introduction}

The implementation of quantum computation using linear optical elements and
measurements with post-selection has attracted a great deal of attention
since the seminal work by Knill, Laflamme, and Milburn~\cite{knill2001a}. 
In that paper the authors showed how the ideas of linear optical manipulation
of photons, together with photodetection and postselection, can be combined
with the concept of state teleportation through a quantum gate~\cite{gottesman1999a}
to perform universal quantum computation. The price paid is that the two-qubit
gates become non-deterministic. Whenever the gate  fails, our qubit is measured in the 
computational basis with known outcome. A very important feature of this scheme is
that gate failures are known, and this can be used to implement error 
correcting codes tailored to this particular situation.

In \cite{knill2001a} the authors also showed that the success probability of the 
controlled-sign gate (CSIGN) can be made arbitrarily close to one by adding more
ancilla modes. This result, combined with the existence of a threshold for
quantum computation~\cite{nielsen2000} implies that only a constant overhead is required
to implement gates whose failure probability is below the threshold. A naive calculation
shows that of the order of $10^4$ ancilla modes are required per two-qubit gate, which is
difficult to achieve in practice. The situation can be improved by devising 
error-correcting codes that exploit characterisitics of the error model. This reduces 
the overhead required, but it still does not render the scheme easy to implement.

An approach that seems to be closer to being practical was proposed by
Nielsen~\cite{nielsen2004a}. Instead of using the ideas of linear-optical
quantum computing (LOQC) to perform a quantum computation in the usual quantum circuit model,
Nielsen proposed using the LOQC CSIGN gate to build optical cluster states that
can then be used to do quantum computations as proposed by 
Raussendorf and Briegel~\cite{raussendorf2001a}. The key point is that higher
probabilities of gate failure can be tolerated while still being able to
construct the required cluster state. Furthermore, even a gate that succeeds with arbitrarily
small probability can be used to efficiently build a cluster 
state~\cite{barrett2005a,duan2005a}. This allows an LOQC gate with only a small number
of ancilla modes to be sufficient for building the cluster state.

In its most basic form, the KLM scheme is extremely fragile against 
detectors errors. Failure to detect a photon introduces errors into
the quantum state that are not detected. To minimize their effect on the quantum
computation we need to have detectors with efficiencies above 99\%, which are far beyond
from what is currently available. A modified scheme was also presented in 
Ref.~\cite{knill2001a}, that would detect this photon loss at the price of
requiring a more complicated entangled ancilla, with double the number
of path modes required by the basic scheme. This would considerably complicate
the implementation of this loss-detecting gate since the associated 
interferometer would be more difficult to control and stabilize. 
Nevertheless, such loss-detection mechanism is crucial even for the application
of the basic KLM scheme to the construction of optical cluster states, since imperfect
detectors can significantly affect the performance of the gate, introducing errors
with probability 
of 30\% and higher for currently available
detectors, \emph{even when the
gate is assumed to be successful}.  

The implementation of the KLM scheme using polarization encoding has two
useful features. First, it requires half the number of path modes to implement
the gates, which makes the associated interferometers easier to setup and
control. And second, when using polarization encoding, the basic form of the KLM
scheme \emph{already has a photon loss detection mechanism}. The 
underlying reason for this feature emerges in a more natural way than 
for the modified dual-rail KLM scheme, and it is due to the conservation of
the number of photons passing through a linear optical setup composed of
mirrors, beam-splitters and phase-shifters. Even with the photon loss
detection mechanism, the KLM scheme requires high-efficiency single-photon
sources and high-efficiency detectors to apply the non-linear phase gate
in the construction of an entangled ancilla with the required high-fidelity.
By implementing the scheme in the polarization basis, we can completely discard the
requirement for high-efficiency detectors, if we replace the single-photon sources by
high-fidelity Bell-pair sources. The errors introduced by the gate will then be
due only to the dark-count rate of the detectors, which can be as low as $10^{-8}$
dark counts per gate. Thus, successful gates will also be high-fidelity gates, and
they can be used to construct high-fidelity optical cluster states.

This paper is organized as follows. In Section II we present a detailed calculation of
the KLM scheme using polarization encoding, including the application of the
CSIGN gate and the generation of the required entangled ancillas. In Section III
we discuss the effect of errors due to non-ideal detectors, both in the original
KLM scheme and the one with polarization encoding, and show that the loss detection
mechanism is crucial for the construction of a high-fidelity non-deterministic
gate. In Section IV we discuss
applications to the construction of optical cluster states and finally in Section V
we present our conclusions.

\section{The KLM scheme with polarization encoding}

One of the pillars on which the KLM scheme is based is the near-deterministic 
teleportation
of the state of an optical mode using linear-optical elements, photodetectors, and an
ancilla prepared in a particular entangled state. The success probability of this
teleportation depends on the number of ancilla modes. For an ancilla state of $2n$ modes,
the probability of success is $\frac{n}{n+1}$. The teleportation procedure goes as
follows. First, the mode to be teleported together with the first $n$ ancilla modes
are sent through an optical device that performs a Fourier transform among the modes.
This device can be constructed using beam-splitters, phase shifters, and mirrors.
After the Fourier transform, we measure the number of photons present in each mode.
For this step, number-resolving photodetectors are required. If the total number 
of photons measured is $0$ or $n+1$, the gate failed and our qubit is measured in the 
computational basis. If $k$ photons are measured, with $0<k<n+1$, then the state of
our qubit can be recovered by post-selecting mode $n+k$  of the ancilla and applying
a phase shift that depends on the distribution of the photons measured among
the first $n+1$ modes. The same can be accomplished with polarization encoding
as we show below.

\subsection{Near-deterministic teleportation}

We will encode the state of our qubit (corresponding to an optical mode) into the
polarization of the photon, so we will have $|0\rangle \rightarrow |H\rangle$ and
 $|1\rangle \rightarrow |V\rangle$. The state of our qubit will then be written as
\begin{equation}
\label{psi}
|\psi\rangle = \alpha |H\rangle +\beta |V\rangle.
\end{equation}
The ancilla state required to implement the teleportation has the same form as the
state used in the original KLM paper with $0$ replaced by $H$ and $1$ by $V$. We will keep
the same notation used in Ref.~\cite{knill2001a} and note this state by $|t_n \rangle$.
Then we have
\be
\label{tele}
|t_n \rangle = \frac{1}{\sqrt{n+1}} \sum_{j=0}^n 
|V\rangle^j |H\rangle^{n-j} |H\rangle^j |V\rangle^{n-j}.
\ee     
The main difference between this teleporting state and the one used in the original KLM
is that (\ref{tele}) has exactly one photon per mode, for a total of $2n$ photons.
The analogous state in the original KLM scheme is 
\be
\label{teleKLM}
|t_n \rangle_{KLM} = \frac{1}{\sqrt{n+1}} \sum_{j=0}^n 
|1\rangle^j |0\rangle^{n-j} |0\rangle^j |1\rangle^{n-j},
\ee 
that has only $n$ photons in $2n$ modes. This difference will allow us to detect
photodetector failure and hence minimize the errors introduced by the gate.

We will now give a detailed calculation of how the teleportation works. It is 
useful to write the states in terms of creation operators applied to the vacuum.
We will call $a_k^\dagger$ the
creation operator of a \emph{vertically} polarized photon in mode $k$, and
$b_k^\dagger$ the creation operator of a \emph{horizontally} polarized photon
in mode $k$. Then we have 
\bea
a_k^\dagger |vac\rangle & = & |V\rangle_k \nonumber \\
b_k^\dagger |vac\rangle & = & |H\rangle_k,
\eea
where $|vac\rangle$ represents the vacuum state. We will write
$|vac\rangle_{1\ldots n}$ to represent the vaccum state of modes
$1$ to $n$.  

The Fourier transform applied to a set of $n+1$ modes, that
for convenience we will call modes $0$ to $n$, is given in terms
of its action on the creation operators
\bea
\label{fourier}
\hat{F_n} (a_k^\dagger) & = & \frac{1}{\sqrt{n+1}} \sum_{l_k = 0}^n
\omega^{k l_k} a_{l_k}^\dagger \nonumber \\
\hat{F_n} (b_k^\dagger) & = & \frac{1}{\sqrt{n+1}} \sum_{l_k = 0}^n
\omega^{k l_k} b_{l_k}^\dagger,
\eea
where $\omega = e^{i \frac{2 \pi}{n+1}}$. This Fourier transform can
be implemented with linear optical elements~\cite{reck1994a}. One important 
point is that this operation does
not mix the polarizations of the photon, which can be seen from 
Eqs. (\ref{fourier}) in the fact that the creation operators for
each polarization transform among themselves.

We can now rewrite the teleporting state $|t_n\rangle$ using 
the creation operators for horizontally and vertically polarized
photons. So, we get
\begin{widetext}
\be
|t_n\rangle = \frac{1}{\sqrt{n+1}} \sum_{j=0}^n a_{1}^\dagger
\ldots a_{j}^\dagger b_{j+1}^\dagger \ldots b_{n}^\dagger
b_{n+1}^\dagger \ldots b_{n+j}^\dagger a_{n+j+1}^\dagger \ldots
a_{2n}^\dagger |vac\rangle_{1\ldots 2n}.
\ee
The teleportation trick starts by considering the joint state formed
by our qubit in state (\ref{psi}) together with the state $|t_n\rangle$.
Expanding this, we have
\be
\label{psitn}
|\psi\rangle|t_n\rangle  =  \frac{1}{\sqrt{n+1}} \sum_{j=0}^n
\left\{ \alpha b_0^\dagger \left( \prod_{k=1}^j a_k^\dagger \right)
\left( \prod_{k=1}^n b_{j+k}^\dagger \right)
\left( \prod_{k=1}^{n-j} a_{n+j+k}^\dagger \right) 
 + \beta a_0^\dagger \left( \prod_{k=1}^j a_k^\dagger \right)
\left( \prod_{k=1}^n b_{j+k}^\dagger \right)
\left( \prod_{k=1}^{n-j} a_{n+j+k}^\dagger \right) \right\} 
|vac\rangle_{0\ldots 2n}.
\ee
\end{widetext}
This is a state of $2n+1$ modes. Note that the difference between
the two terms is, besides the values of $\alpha$ and $\beta$, that
the first term has a creation operator for a horizontally polarized
photon in mode $0$, while the second has a creation operator
for a vertically polarized photon in that mode.

The next step is to apply the
Fourier tansform to the first $n+1$ modes (i.e., modes $0$ to $n$). Note that,
since the two terms have different numbers of creation operators of each type
($H$ or $V$), and since the Fourier transform does not mix polarizations,
the same will hold after the transformation is applied. The state obtained after 
the Fourier transform is
\begin{widetext}
\be
\label{aftfou} 
\left( \frac{1}{\sqrt{n+1}}\right)^{n+2}  \sum_{j=0}^n
\left\{\sum_{0\leq l_0 ,\ldots,l_n \leq n} \omega^{\sum_{k=0}^{n} k l_k}
\left(\alpha b_{l_0}^\dagger a_{l_1}^\dagger \ldots a_{l_j}^\dagger 
b_{l_{j+1}}^\dagger \ldots b_{l_n}^\dagger 
+ \beta a_{l_0}^\dagger a_{l_1}^\dagger \ldots a_{l_j}^\dagger 
b_{l_{j+1}}^\dagger \ldots b_{l_n}^\dagger \right) |vac\rangle_{0\ldots n}
\right\} \underbrace{|H\rangle^j |V\rangle^{n-j}}_{modes (n+1,\ldots, 2n)}.
\ee
\end{widetext}
Note that the $\alpha$ terms have $j$ of the V-photons and $n-j+1$ of the H-photons,
while the $\beta$ terms have $j+1$ of the V-photons and $n-j$ of the H-photons.
This difference will be responsible for transfering the superposition in
the state of mode $0$ to one of the last $n$ modes of the ancilla. 

Now the idea (following KLM) is to perform a measurement that collapses
the state vector (\ref{aftfou}) to a certain value of $j$ for the 
$\alpha$ terms, and to $j-1$ for the $\beta$ terms. In the KLM scheme
this is accomplished by measuring the number of photons in each
of the $n+1$ output modes of the Fourier transform. Here however, that is not
enough since there are two kinds of photons, so just measuring the number
of photons present in each mode \emph{ does not collapses the
state vector (\ref{aftfou}) into the state we want}. To solve this problem
we just need to perform a stronger measurement that tells us not just
how many photons are in one mode \emph{but how many of each
polarization are present}. This measurement can be easily implemented
by sending the output of each mode through a polarizing 
beam-splitter (PBS), and then measuring the number of photons present
in each of the two output ports of the PBS. This requires the same
number-resolving photodetectors used for the original KLM scheme, with the only
difference that we now need twice as many.

Let us assume that we have performed this measurement, and we
have obtained that in mode $j$, there are $r_j$ of the V-photons and
$h_j$ of the H-photons. Note that since there was one photon per mode in
the first $n+1$ modes of (\ref{psitn}), the total number of photons
measured at the output of the Fourier transform \emph{must be $n+1$}.
This is the key feature that allows us to know when a detector fails to detect
a photon. But now we want to know what is the state of the whole system after
this projective measurement. First, let us consider the two simplest cases.
If $\sum_{j=0}^n r_j = n+1$, then all the photons detected are V-photons
(i.e., $h_j = 0,\ \forall j$). Looking at (\ref{aftfou}), we can see that the
only term that has $n+1$ of the V-photons in the first $n+1$ modes corresponds
to the $\beta$ term with $j=n$. Any other term in (\ref{aftfou}) has at least
one H-photon. Then the state corresponding to that measurement result is to be 
\be
|V\rangle^{n+1} |H\rangle^n.
\ee
This correponds to a projective measurement of our qubit in the computational
basis. The superposition is destroyed and the teleportation failed. 
The probability of obtaining this measurement result is  $\frac{|\beta |^2}{(n+1)}$.
Similarly, if we measure that $\sum_{j=0}^n r_j = 0$, that means
$\sum_{j=0}^n h_j = n+1$, and we can repeat the reasoning above
replacing V-photons by H-photons. So again, the result is a projective
measurement in the computational basis, which destroys the superposition.
The probability of this event ocurring is $\frac{|\alpha |^2}{(n+1)}$, and
so the total probability of failure of the teleportation is $\frac{1}{(n+1)}$ independent
of the input state.

So let us now assume that $\sum_{j=0}^n r_j \neq n+1,0$, and write
$\sum_{j=0}^n r_j = k$. Then we also have $\sum_{j=0}^n h_j = n-k+1$,
since the total number of photons detected is always $n+1$. The state 
corresponding to that measurement result is  
\begin{widetext}
\bea
\label{aftmeas}
\left(\frac{1}{\sqrt{n+1}}\right)^{n+2} & \left\{ 
\sum_{\cal S}  \omega^{\sum_{p=0}^{n} p\, l_p}
\alpha b_{l_0}^\dagger a_{l_1}^\dagger \ldots a_{l_k}^\dagger 
b_{l_{k+1}}^\dagger \ldots b_{l_n}^\dagger |vac\rangle_{0\ldots n}
|H\rangle^k |V\rangle^{n-k} + \right.  \nonumber \\
& \left.  +  \sum_{\cal S'}  \omega^{\sum_{p=0}^{n} p\, l_p} 
\beta a_{l_0}^\dagger a_{l_1}^\dagger \ldots a_{l_{k-1}}^\dagger 
b_{l_{k}}^\dagger \ldots b_{l_n}^\dagger  |vac\rangle_{0\ldots n}
|H\rangle^{k-1} |V\rangle^{n-k+1} \right\},
\eea
with 
\bea
{\cal S}  = & \left\{ (l_0,\ldots,l_n) / \right. & \{l_1,\ldots,l_k \} 
\,\mathrm{contains \
the \ value}\ j, r_j \,\mathrm{times}, \, \mathrm{and} \nonumber \\
 & & \left. \{l_0,l_{k+1},\ldots,l_n\}\, \mathrm{contains\ 
the \ value}\ j, h_j \,\mathrm{times},\, j \in \{0,\ldots,n\} \right\} \nonumber \\
{\cal S'}  = & \left\{ (l_0,\ldots,l_n) / \right. & \{l_0,\ldots,l_{k-1}\}\, 
\mathrm{contains\ 
the \ value}\ j, r_j \,\mathrm{times}, \, \mathrm{and}  \nonumber \\
 & & \left. \{l_{k},\ldots,l_n\} \, \mathrm{contains \ 
the \ value}\ j, h_j \,\mathrm{times},\, j \in \{0,\ldots,n\} \right\}.
\eea

\end{widetext}

By looking at the two sums in (\ref{aftmeas}) we can see that these
two terms have the same state for the first $n+1$ modes since they have the
same number of V-photons and H-photons in each of the first $n+1$ modes (fixed by the 
result of the measurement). The only difference is given by the state 
of the last $n$ modes and by the factors introduced by the two sums
\be
\label{sums}
 \sum_{\cal S}  \omega^{\sum_{p=0}^{n} p\, l_p} \ \ \mathrm{and}
 \ \ \ \sum_{\cal S'}  \omega^{\sum_{p=0}^{n} p\, l_p}.
\ee
Since the sums are over two different sets of $(n+1)$-tuples, it is
not clear if this will change the relative weights in the 
superposition given by $\alpha$ and $\beta$. However, after a little algebra
it can be shown that these two factors differ only by an overall phase. More
precisely we have
\be
\sum_{\cal S'}  \omega^{\sum_{p=0}^{n} p\, l'_p} =
\sum_{\cal S}  \omega^{\sum_{p=0}^{n} p\, l_{p}}\
\omega^{-\sum_{p=0}^{n}\, l_{p}},
\ee
where $\omega = e^{\frac{2 \pi i}{n+1}}$ as defined earlier. By following the
above calculation carefully, it is not hard to show that 
\be
\sum_{p=0}^{n} l_{p} = \sum_{j=0}^n j (r_j + h_j).
\ee
It is worth noting that since  $(r_j +h_j)$ is the total number of photons measured
in mode $j$, this exponent has exactly the same form as the dephasing introduced
by the teleportation procedure 
in the original KLM scheme. In summary we have that
\be
\sum_{\cal S'}  \omega^{\sum_{p=0}^{n} p\, l'_p} =
\omega^{-\sum_{j=0}^n j (r_j + h_j)}\ 
\sum_{\cal S}  \omega^{\sum_{p=0}^{n} p\, l_{p}}.
\ee
Taking all of this into account we can rewrite the state 
(\ref{aftmeas}) as
\be
|\Phi\rangle_{0\ldots n} |H\rangle^{k-1} 
\left(\alpha |H\rangle + \beta \, \omega^{-\sum_{j=0}^n j (r_j + h_j)}
|V\rangle \right) |V\rangle^{n-k+1},
\ee
where $|\Phi\rangle_{0\ldots n}$ is a normalized state of the first $n+1$ modes
fixed by the result of the measurement. We can see that the superposition
was teleported to the mode $n+k$, up to a known relative phase. But since
we know exactly the value of that phase, we can get rid of it by 
using phase shifters, and then the final state becomes
\be
\label{final}
|\Phi\rangle_{0\ldots n} |H\rangle^{k-1} 
\left(\alpha |H\rangle + \beta 
|V\rangle \right) |V\rangle^{n-k+1}.
\ee  
The last $n$ modes, with the exception of mode $n+k$ of course, are
left in a known state and can be reused later. It is worth noting that 
the state (\ref{final}) \emph{has exactly the same form} as the state that we obtain 
in the original KLM scheme, if we replace all H's by zeros and all V's by ones.

\subsection{Near-deterministic CSIGN}

The near-deterministic teleportation procedure described above can be 
combined with the idea of applying a quantum gate through teleportation
to perform a near-deterministic CSIGN gate. To do this we will need 
a teleporting state of $4n$ modes, which is nothing but two copies of
the teleporting state $|t_n \rangle$ with CSIGN gates applied to
the mode pairs $(n+k,3n+l)$, $1\leq k,l\leq n$. This (unnormalized) 
state can be written as
\begin{widetext}
\be
|t_n'\rangle = \sum_{i,j=0}^n (-1)^{(n-j)(n-i)} |V\rangle^j 
|H\rangle^{n-j} |H\rangle^j |V\rangle^{n-j} \times
|V\rangle^i 
|H\rangle^{n-i} |H\rangle^i |V\rangle^{n-i}.
\ee
\end{widetext}
Note that this is a state of $4n$ modes that has $4n$ photons. To apply a CSIGN gate
to two modes $A$ and $B$ we can proceed by teleporting the mode $A$ using the first
$2n$ modes of $|t_n' \rangle$ and then teleporting mode $B$ using the last $2n$ modes
$|t_n' \rangle$. Each teleporting step will proceed as before, 
requiring post-selection
and phase correction, and failing independently with probability $\frac{1}{n+1}$.
The calculation is essentially the one presented in the previous subsection applied
twice, so we will not present it explicitly here.

\subsection{State preparation}

As described in Ref.~\cite{knill2001a} the KLM scheme for LOQC reduces to a state
preparation problem: we need to be able to construct the state $|t'_n \rangle$ using
only linear optics and photodetection. In the original scheme with dual-rail encoding,
we proceed as follows. From the state $|01\rangle |01\rangle$ of four optical
modes (generated using a single photon source), we construct the
state 
\be
\frac{1}{2} (|01\rangle + |10\rangle)(|01\rangle + |10\rangle), 
\ee
by applying beam splitters to the mode pairs $(1,2)$ and $(3,4)$. Then we send modes
$1$ and $3$ through a Mach-Zender interferometer with a (non-deterministic) non-linear 
sign-shift gate applied to each arm. This gate (which can be performed using linear optics
and photo-detection) applies the transformation
\be
\alpha_0 |0\rangle + \alpha_1 |1\rangle +\alpha_2 |2\rangle
\rightarrow \alpha_0 |0\rangle + \alpha_1 |1\rangle -\alpha_2 |2\rangle.
\ee
After the interferometer the state of the system is
\be
|t'_1 \rangle_{KLM} = \frac{1}{2}(|1010\rangle + |0110\rangle + |1001\rangle - |0101\rangle),
\ee
which can be used to apply a CSIGN gate with probability $\frac{1}{4}$. To apply this gate with 
higher success probability we need to construct the states $|t'_n \rangle$ with $n>1$. 
This is done by a recursive procedure that uses $|t'_p \rangle$ to build $|t'_n \rangle$,
where $p<n$. By recycling these resources whenever a gate fails, it can be shown that
the number of trials required to build $|t'_n \rangle$ scales as $2^{O(\sqrt{n})}$. 

Preparing the corresponding states with polarization encoding can be done much in the
same way with some minor but essential modifications. First we should note that 
the analog of the beam-splitter transformation with dual-rail encoding, which
takes the form
\be 
|01\rangle \rightarrow \cos \theta\, |01\rangle + \sin \theta \,|10 \rangle,
\ee
cannot be implemented with linear optics if we are using polarization encoding, 
since in this
case it takes the form
\be 
|HV\rangle \rightarrow \cos \theta\, |HV\rangle + \sin \theta\, |VH \rangle,
\ee
which is an entangling operation between two photons and cannot be implemented 
with linear optics alone. However, this operation 
can be applied non-deterministically if we are allowed to use the
non-deterministic CSIGN. To see this, first note that the required trasformation is
represented by the unitary matrix (in the basis $\{ |HH\rangle, |HV\rangle, |VH\rangle,
|VV\rangle \}$)
\be
\left( \begin{array}{cccc}
        1 & 0 & 0 & 0 \\
        0 & \cos {\theta} & \sin {\theta} & 0 \\
        0 & -\sin {\theta} & \cos {\theta} & 0 \\
        0 & 0 & 0 & 1 
       \end{array}
\right),
\ee
and this matrix can be implemented by the quantum circuit
\begin{figure}[ht]
\centerline{\includegraphics[scale=0.46]{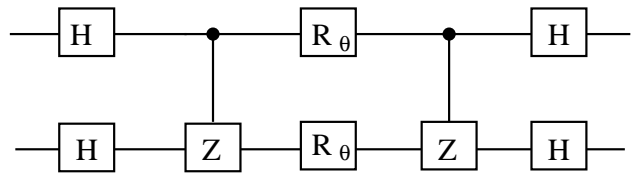}}
\caption{\label{HVBS} Quantum circuit that applies the HV-beam splitter.}
\end{figure} 
All the one qubit operations can be applied deterministically, while the CSIGN (or C-Z)
can be applied with some probability. 

Let us assume for the moment that we can construct the state $|t'_1 \rangle$ with 
polarization encoding. Then we can generate the states $|t'_n \rangle$ for $n>1$ 
following the same algorithm used in the original KLM scheme if we replace
all beam-splitters by the circuit of Fig. \ref{HVBS}. Following that procedure
it is not hard to see that the only effect is that the number of non-deterministic
gates required (roughly) doubles, but it remains linear on $n$. Thus the resource
required to build the teleporting states $|t'_n \rangle$ with polarization 
encoding are only polynomially bigger than the resources required in the original
KLM scheme.

Finally, we need to show that we can construct the state 
\be
\label{tp1pol}
|t'_1 \rangle = \frac{1}{2} (|VHVH\rangle + |HVVH\rangle + |VHHV\rangle - |HVHV\rangle),
\ee 
using only linear optics and photodetection. It is interesting to note that this
state is the same four-photon cluster state used by Walther \emph{et al.}
in their experimental demonstration of a one-way quantum computer~\cite{walther2005a}. 
First we note that we can deterministically 
transform a state with dual-rail encoding into a state with polarization encoding
by using a polarizing beam-splitter (PBS), so we have
\be
\alpha |10\rangle + \beta |01\rangle \rightarrow \alpha |H\rangle + \beta |V\rangle.
\ee
Now we can apply the quantum encoder introduced by Pittman {\it et al.}~\cite{pittman2001a}
plus polarization rotators 
to perform, with probability $\frac{1}{2}$, the operation
\be 
\alpha |H\rangle + \beta |V\rangle \rightarrow \alpha |VH\rangle + \beta |HV\rangle.
\ee
Composing these two transformations we can perform
\be 
\label{transf}
\alpha |10\rangle + \beta |01\rangle \rightarrow \alpha |VH\rangle + \beta |HV\rangle,
\ee
with probability $\frac{1}{2}$.
Now consider the state
\be
\label{tp1klm}
\frac{1}{2}(|1010\rangle + |0110\rangle + |1001\rangle - |0101\rangle).
\ee
This is nothing but the state $|t'_1 \rangle_{KLM} $ in the original KLM scheme, and we know
it can be constructed with probability $\frac{1}{16}$ using linear optics and photodetection.
By applying the transformation (\ref{transf}) to the mode pairs $(1,2)$ and $(3,4)$
in (\ref{tp1klm}),
we obtain the state (\ref{tp1pol}) with probability $\frac{1}{64}$. This probability 
is not claimed to be optimal.

\section{Errors introduced by real detectors}

Both the original KLM scheme and its implementation with polarized photons described
above assume all the detectors are perfect, i.e, they have unit quantum efficiency
and zero dark counts. If we allow for real detectors and still require the scheme
to apply the entangling gates with a probability of failure low enough such that
error correction and fault tolerant design allows for arbitrarily long quantum
computation, the required efficiencies turn out to be extremely high, compared to the
presently available detectors. A naive estimation requires a detector efficiency
higher than 99.99\%. A clever use of error correction, exploting the properties
of the error model, may reduce this requirement. However, no significant improvement
that would render the scheme viable with currently available detectors has been
proposed so far. 

On top of the high-efficiency requirement for the detectors, the scheme requires
a rather large overhead in terms of extra ancilla modes. Again, a naive calculation
of this overhead yields $10^4$ ancilla modes required per CSIGN gate. Exploiting
the properties of the failure modes can reduce this requirement to about $50$ ancilla
modes per CSIGN gate. This is still a rather large number to be practical for
an actual implementation.

A big step forward in reducing this required overhead was the proposal by
Nielsen~\cite{nielsen2004a} to combine the techniques of the KLM scheme
with the cluster-state model of quantum computation~\cite{raussendorf2001a}. Rather
than using the CSIGN gate for the actual computation, we can use it to construct
a cluster state on which we can later perform our quantum computation. The advantage
of this approach is that lower success probabilities can be tolerated with only
a modest overhead on the time required to build the cluster state, thus reducing the
extra ancilla modes to four. 

However, this proposal cannot escape the requirement of very high efficiency detectors.
The reason behind this is the fact that low-efficiency detectors will make the
fidelity of the CSIGN gate low \emph{even when the gate is assumed to be successful}.
This will reduce the fidelity of the cluster state and require a larger state to
accomodate for fault tolerance and error correction during the cluster state computation.
At this point, using polarization encoding instead of dual-rail encoding shows a clear
advantage. Polarization encoding allows for \emph{very high fidelity CSIGN gates},
conditioned on gate success (at the price of a lower probability of success), even
with currently available detectors. Since a cluster state can be efficiently
constructed using gates with arbitrarily small success probability~\cite{barrett2005a,
duan2005a}, using polarization encoding allows for efficient construction of
high-fidelity cluster states. In the remainder of this section, we will analyze
in more detail the errors introduced by real detectors for both the original
KLM scheme and the one with polarization encoding, and discuss the advantages of the latter.

\subsection{Errors in KLM}

We will model a real detector with two parameters: a quantum efficiency $\eta$ and
a dark count rate $\lambda$. We will assume that the dark counts follow a Poisson
distribution, so the probability of having $d$ dark counts during the measuring
interval $\tau$ will be
\be 
D(d) = e^{-\lambda \tau} \frac{(\lambda \tau)^d}{d!}.
\ee
We can then write the conditional probability of the detector measuring $k$ photons
when $l$ photons were present as~\cite{lee2004a}
\be
\label{pdetcond}
P_D (k|l) = \sum_{d=0}^k D(k-d) {l \choose d} \eta^d (1-\eta)^{l-d}.
\ee

We now analyze the effects of real detectors on the implementation of
the near-deterministic teleportation step, which is the basis of the
whole scheme. Let us recall that, to perform that step, we need to measure
the number of photons present in each of the first $n+1$ modes after applying 
a Fourier transform to them. The total number of photons measured, $k$, tells us
whether the gate succeeded (if $k\ne 0,n+1$), and in that case to which one of
the last $n$ modes of the ancilla was the state of our qubit teleported to. 
It is then clear that accurately determining the value of $k$ is essential 
for the success of the scheme. Imperfect detectors will degrade our ability to
determine $k$.

The finite efficiency and dark-count rate of the detectors will 
sometimes produce a result $k'$ for the total number of photons
that is different from the actual number of photons present $k$, since
some detectors may fail to detect one or more photons, while others
will register dark counts. (For actual detectors, failure to detect a photon 
has a higher probability than dark counts, but we will keep our analysis general.) 
If $k' \ne 0,n+1$, we will assume
that the teleportation was successful and \emph{post-select the wrong
output mode as the one carrying the state of our qubit}. This wrong mode
will be in either the $|0\rangle$ state or the $|1\rangle$ state, depending
on whether $k'<k$ or $k'>k$. This is similar to the measurement error introduced
by the teleportation failure associated with $k=0,n+1$ in the original scheme.
The difference is that \emph{this new failure goes completely undetected}. 
The gate is assumed to have succeeded when in fact it has introduced a measurement
error that will propagate through the computation. 

Another undetected error may occur when the number of non-detected photons is equal to
the number of dark counts. In this case the total number of photons is
correct, but their distribution among the first $n+1$ modes may have been
changed. Since this distribution determines the phase correction that needs to be
applied after a successful teleportation, a phase error may be introduced in
the computation.

As we can appreciate, considering real detectors can modify the KLM scheme
significantly. The main change is that the probability of a detected failure $p_f$
is no longer $1-p_s$, where $p_s$ is the probability of success, since we now
have \emph{non-detected errors} ocurring with some probability $p_{nde}$. 
Also, the probability of a detected failure may no longer be independent of
the input state. Actually, it depends on the input state whenever the probability
of dark counts is different from the probability of photon non-detection, as
is the case for currently available detectors. The root of this is our encoding
of information in photon number, while the failure of the detectors is biased
towards decreasing photons numbers (dark counts are usually negligible with respect to
detection failure). 

Using the conditional probabilities given by (\ref{pdetcond}) we can compute the
probabilities of detected failure $p_f$, and the probability of errors
introduced by the teleportation conditioned on no detected errors, defined
by $p_e = p_{nde}/(1-p_f)$ for different values of detector efficiency
and dark-count rates as a function of the number of ancilla modes $n$.
In particular, we considered the parameters corresponding to the 
number-resolving photo-detector demonstrated by Miller {\it et al.}~\cite{miller2003a}.
In that work a 20\% quantum efficiency was reported, with dark counts
of the order of $10^{-7}$, when considering a measuring time $\tau \sim 100 \mu s$.
It was also reported that this scheme could be improved to achieve an efficiency of 80\%.
Using this value, we plotted the different probabilities for $n=2,3,4$. 
\begin{figure}[ht]
\centerline{\includegraphics[scale=0.8]{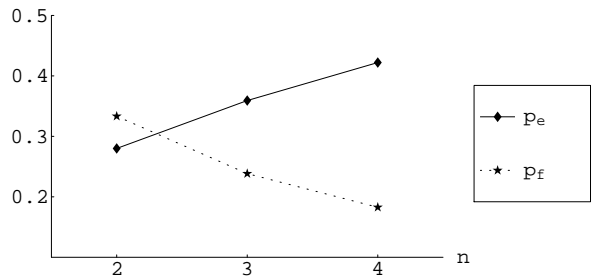}}
\caption{\label{pklm08} Probability of detected failure $p_f$ and probability of error
$p_e$ introduced by a ``successful'' gate in the KLM scheme ($\eta = 0.8$ , $
\lambda \tau \sim 10^{-7}$).}
\end{figure}
As we can see from Fig. \ref{pklm08}, the probability of a detected failure still 
decreases with $n$. However, the probability of an error being introduced
by the gate when no failure is detected increases with $n$. Furthermore,
this error is rather high ($\sim 27\%$) even for $n$ as small as $2$.
For $n=4$ this error is greater than $40\%$. This can be easily understood
since increasing $n$ increases the number of non-detected failure modes.
From this point of view, increasing $n$ is actually counterproductive. 

\subsection{Errors with polarization encoding}

The effect of errors introduced by real detectors when we use polarization
encoding is remarkably different. As discussed before, the role of the total number
of photons is played by the total number of vertically polarized photons.
Since both the teleporting state and the input qubit have exactly one photon
per mode, the total number of photons (both vertically and horizontally polarized)
measured after the Fourier transform \emph{must be $n+1$}. If this number is different 
from $n+1$ then we know for sure that some of the detectors failed, and we have
lost the information about which mode the input qubit was teleported to. This
case should be considered as a detected failure of the gate. Note that the most common
error for real detectors (failure to detect a photon) will be recognized, while it
would have gone undetected in the original KLM scheme. 

\begin{figure}[ht]
\centerline{\includegraphics[scale=0.7]{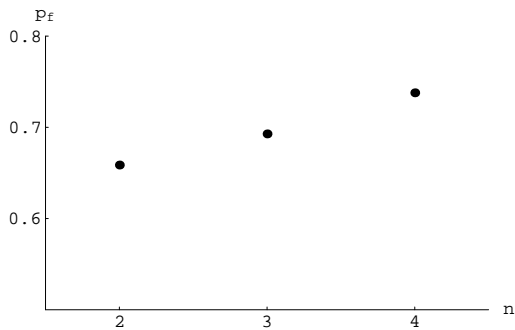}}
\caption{\label{pfp} Probability of detected failure $p_f$ using polarization encoding 
($\eta = 0.8$ , $
\lambda \tau \sim 10^{-7}$).}
\end{figure}

Even if we use polarization encoding, there would be undetected errors occurring
when we assume the gate has succeeded. These errors will require the same number 
of detector
failures to measure a photon and registered dark counts. Depending on which
detectors register these failures, we may choose the wrong output mode (i.e., we
have the wrong information about the total number of vertically polarized photons
present), or we may introduce a phase error (total number of V-photons is
correct but their distribution among modes is not). The key point is that 
the probability of these type of errors is dramatically reduced because of the
low probability of dark counts in currently available detectors. Using the same
detector parameters discussed above, we computed the probability of detected failure
$p_f$ for the teleportation step (see Fig. \ref{pfp}).
We can see that the probability of a detected failure is greater than for the
KLM scheme ($66\%$ compared to $33\%$ for $n=2$). Furthermore, this probability increases
with $n$ instead of decreasing. This can be easily understood. First, the probability
of a detected failure depends essentially on the efficiency of the detectors. When using
polarization encoding, we require double the number of detectors than for the KLM scheme,
thus it is not surprising that our gate has a higher probability of failing. The scaling
with $n$ may not have been foreseen but should not be unexpected, since now the
failure probability includes errors derived from detector failures that are more
abundant for higher values of $n$, as the number of detectors required is $2(n+1)$. 

\begin{figure}[ht]
\label{errorpol}
\centerline{\includegraphics[scale=0.7]{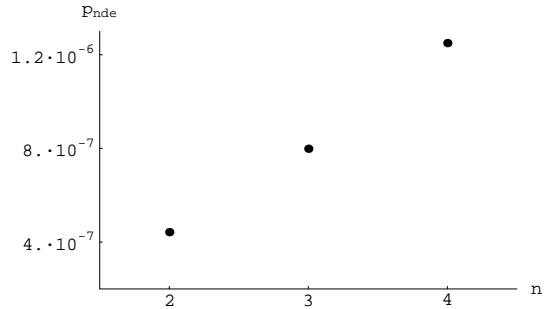}}
\caption{\label{pep} Probability of error $p_e$ introduced by the gate when 
no failure was detected and using polarization encoding 
($\eta = 0.8$ , $
\lambda \tau \sim 10^{-7}$).}
\end{figure}

We have also computed the probability of error conditioned on
no detected failure, $p_e = p_{nde}/(1-p_f)$, with $p_{nde}$ the probability
of a nondetected error. The results are shown in Fig. \ref{pep}.      
This error probability is \emph{several orders of magnitude lower than the corresponding
one for the KLM scheme} using the same detector parameters. Its order of magnitude is
given roughly by the order of magnitude of the dark-count probability since the most 
likely error corresponds to one detector failing to register a photon and
another detector registering a dark count (events with higher number
of dark counts are strongly suppressed). Lowering this probability can be achieved
by reducing the dark count rate, independent of the detector efficiency. We still have
the same behavior with increasing $n$ that is related to the higher number
of events associated with errors as the number of detectors required grows.

\section{Application to cluster state quantum computing}

As we discussed before, coupling the KLM scheme with the cluster state model of quantum 
computation greatly reduces the number of ancilla modes required. This, together 
with the existence
of protocols to efficiently build these cluster states with non-deterministic
gates, makes this approach very appealing.
For optical cluster states however, other issues need to be addressed. 
For the cluster state approach to be successful we need to be able
to deal with the inevitable errors that will occur during its construction
and during the computation itself. Since the computation proceeds by performing
single qubit measurements while the quantum information is propagated along
the cluster by quantum correlations, an efficient measurement procedure is required.
This is not that simple in an optical implementation since currently available
photodetectors do not have very high quantum efficiency. A good detector may have
an efficiency of 80\%, which means that on average one in five photons of the 
cluster will not be detected. This rate of loss of cluster qubits can make
quantum computing impossible. An incremental encoding was proposed by Gilchrist 
{\it et al.}~\cite{gilchrist2005a} to make the computation fault tolerant to photon
loss. Another idea recently proposed by Varnava {\it et al.}~\cite{varnava2005a},
exploits the fact that the cluster state is an eigenstate of a set of stabilizer
operators to infer the results of measurements on photons that fail to be detected
by the photodetectors. This scheme allows us to perform the required measurements
for a quantum computation provided the efficiency of the detectors is above 50\%
(which are currently available).

Errors can also be introduced during the construction of the cluster, which in turn 
may introduce non-Markovian errors during the computation.
Fortunately, as in the case of the quantum circuit model of quantum computing, there is
an error threshold below which a fault tolerant design allows us to perform
an arbitrarily long computation, although the value of this threshold 
is not yet known for the cluster state model~\cite{nielsen2004b,aliferis2005a}. 
However, it is not expected to be significantly smaller than the
threshold for the quantum circuit model since we can recast the cluster state approach
as a quantum circuit. For example, we can consider the qubits in the cluster to be 
a quantum registry initialized in a particular product state, the CSIGN gates
required to build the cluster can be regarded as the gates in the circuit
model, and the measurements that implement the computation on the cluster can be
considered as part of the readout measurements in the circuit model. Assuming perfect
measurements, this shows that a CSIGN gate with error probability
below the quantum circuit threshold should be enough to construct a cluster state
with fidelity high enough to allow quantum computing in the cluster state model.
The typical value quoted for the error threshold in the quantum circuit model is around 
$10^{-4}$, which corresponds to a general error model. A higher threshold
might be possible if we take advantage of the properties of the error model associated
with a particular implementation.

Thus emerges the greatest advantage of using polarization encoding instead of dual-rail 
encoding. Dual-rail encoding has a very high probability of non-detected errors
being introduced when applying a CSIGN gate, and hence when constructing a cluster
state using Nielsen's approach, unless very high efficiency detectors
are used. On the other hand, polarization encoding allows us to reduce
this non-detected errors independently of the efficiency of the detectors. The
only requirement is that the dark-count rate of the detectors be sufficiently
small, which is actually the case for currently available detectors. For example,
a number-resolving photodetector was reported in Ref.~\cite{miller2003a} with 
a $20\%$ efficiency and dark-count rate of the order of $10^{-9}$ for a measuring
time of $1 \mu{\mathrm sec}$. Another number-resolving detector with an efficiency
of $88\%$ has been reported in Ref.~\cite{rosenberg2005a}. Although this detector was
reported to be essentially noise-free, no measured value of the dark-count rate was
given.

There are several recipes to efficiently build a cluster state
using non-deterministic gates~\cite{barrett2005a,duan2005a}, and any
of these approaches can be used to build an optical cluster
state using the techniques described in this paper. The one specific
advantage that we would like to exploit is the low probability of
non-detected errors when using polarization encoding. As we saw in Fig. \ref{pep}
this probability increases with the number ancilla modes used in the application
of the CSIGN gate, so the smallest entangled ancilla state required
(four photons in four modes)
is actually the one that minimizes the error probability.

\begin{figure}[ht]
\centerline{\includegraphics[scale=0.46]{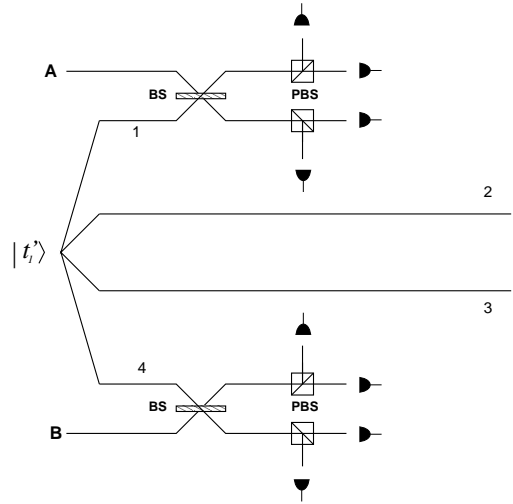}}
\caption{\label{optcircuit} Optical setup required to 
implement a CSIGN between two photons using a four
photon entangled ancilla state. The two input modes A and B are combined through
beam-splitters with modes 1 and 4 of the entangled ancilla $|t'_1\rangle$, respectively. 
After the beam-splitters, the number 
of vertically and horizontally polarized photons in each mode is measured. When the gate
succeeds, modes 2 and 3 carry the state of the input modes with a CSIGN gate
applied to them, up to phase shifts. }
\end{figure}

Using the smallest entangled ancilla state given by
\be
\label{ancilla4}
|t'_1\rangle = \frac{1}{2} (|VHVH\rangle + |HVVH\rangle + |VHHV\rangle - |HVHV\rangle ),
\ee
has some other advantages. In Fig. \ref{optcircuit} we can see the setup required
to implement the CSIGN gate. 
One advantage is that we only need to apply the optical Fourier
transform to a pair of modes and that can be accomplished with a single
balanced beam-splitter. We do not need to apply the more complicated interferometer
required when more than two modes are input to the Fourier transform, which
is experimentally very challenging. Also, since after the measurements
only two modes are left, we do not have to physically post-select the output
modes and only a phase shift may need to be applied to them. 
These two properties could be very useful if we want to integrate this
gate into an optical chip, since no complex interferometers are involved and the
output modes are fixed.

Furthermore, the success or failure of the gate is heralded by the number
of vertically and horizontally polarized photons detected, while the total
number of photons that must be detected is fixed and equal to four. 
Each step of the teleportation
succeeds only when one vertically polarized photon and one horizontally
polarized photon are detected, which in the setup presented in Fig. \ref{optcircuit}
means that two independent detectors must fire. This feature allows us to
implement the same gate with \emph{detectors that are not number-resolving}.
Since a success corresponds to two independent detectors clicking,
we do not need to know how many photons each detector measured. Again,
the only possible errors introduced by the detectors malfunctioning are
related to dark counts, which have an extremely low probability. 
 
Besides the errors introduced by the detectors, the fidelity of this CSIGN gate
can be affected by errors in the entangled ancilla state. If the
state we use is not exactly the state (\ref{ancilla4}), the gate applied will not be exactly
a CSIGN gate, even when the measurements tell us it was successful. 
Thus, to assure the high fidelity of the gate we need to require
the fidelity of the entangled ancilla with respect to the ideal state given by 
(\ref{ancilla4}) to be
high. Since we want the gate to work with fidelity at least $1-10^{-4}$
(to be below the error threshold) we need the fidelity between the two
states to be of the same order. In Section II.C. we showed that this state can be
constructed with linear optical elements and photodetectors, but this construction is
based in the nonlinear sign (NS) gate that is the basis of the KLM scheme,
and the fidelity of the NS gate is very fragile against detector inefficiencies because
it operates in the photon number basis.

\begin{figure}[ht]
\centerline{\includegraphics[scale=0.6]{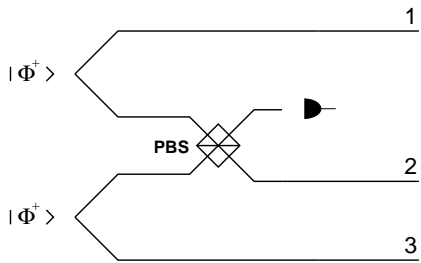}}
\caption{\label{belltoghz} Generation of a polarization GHZ state from two Bell pairs 
using a polarization beam-splitter and photodetection. A detector failure will introduce
an error only in mode 2, which in that case will contain no photons.}
\end{figure}

Nevertheless, we can non-deterministically construct a high efficiency copy of
(\ref{ancilla4}) with linear optics and non-ideal detectors if we have access to
high-fidelity polarization Bell pairs. We first combine two Bell pairs to form 
a GHZ state that might not have high-efficiency. To do this, we mix one 
mode from each Bell pair on a balanced beam-splitter and measure the number
of photons in one of the outgoing modes \emph{irrespective to their polarization},
as seen in Fig. \ref{belltoghz}.
If only one photon is measured by the detector, the state of the remaining three
modes is a GHZ state in the polarization basis. If that number is zero or two, the
procedure is aborted. If non-ideal detectors are used an error may be introduced
when two photons arrive at the detector but only one is registered due to its
non-unit efficiency. This operation is very similar to the type-I fusion
introduced by Browne and Rudolph~\cite{browne2005a}. 
The important point is that even when this error occurs, the only 
output mode affected is the one coming out of the beam-splitter that will
then have no photons. The other two output modes will still have one photon each.

\begin{figure}[ht]
\centerline{\includegraphics[scale=0.6]{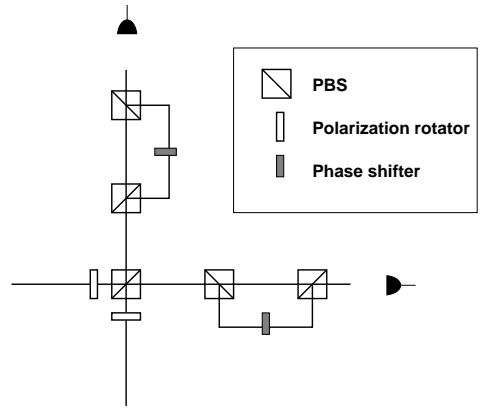}}
\caption{\label{ghztocluster} The two modes of two pairs
of GHZ states constructed according to the scheme in Fig. \ref{belltoghz} that may
have an error, are
sent through the setup shown above. The polarization beam-splitters reflect 
vertically polarized photons. The operation is successful when exactly one photon is
registered by each detector. Then, the state of the remaining four modes of the
the two GHZ pairs can be transformed into $|t_1'\rangle$ using phase-shifters and
polarization rotators.}
\end{figure}

The second step is to take two copies of this GHZ state and combine the two
modes that may have an error (i.e., no photon) using polarization rotators, polarization 
beam-splitters and phase shifters, and then measure the number of photons in each
mode \emph{irrespective to their polarization}, as seen in Fig. (\ref{ghztocluster}).
If there were no errors in the construction of the two GHZ states, we expect exactly two
photons to be detected. If we detect less than two photons, the GHZ states had errors
and the procedure is aborted. If we detect two photons in one of the detectors (and zero 
in the other one), the output state is equivalent to two Bell pairs that can be
reused to generate more GHZ states. If we detect one photon in each mode, the
state of the remaining output modes is given by
\bea
& & \frac{1}{2} (e^{i \frac{\pi}{4}}|HHHH\rangle + e^{-i \frac{\pi}{4}}|HHVV\rangle +
\nonumber \\
& & \ \ \ \ \ \ \ \ \ \ 
+  e^{-i \frac{\pi}{4}}|VVHH\rangle +e^{i \frac{\pi}{4}}|VVVV\rangle ).
\eea
By applying polarization rotators and polarization-dependent phase-shifters, we can
transform this state into (\ref{ancilla4}). The key feature of this construction is that
even though an error may be introduced in the first step by a non-ideal detector, the 
effect of this error will be restricted to a photon missing in only a pair of modes, and it
can be detected when these two modes are mixed and the total number of photons 
is measured~\cite{ralphpriv}. 
Then, an error in the final cluster state can only be related
to the occurrence of a dark count in one of the detectors, and for currently 
available detectors the probability of such event is below $10^{-6}$. As long as the fidelity
of the initial bell pairs is high enough, this procedure generates a four-photon
high-fidelity cluster state that can be used to apply a high-fidelity CSIGN gate.

\section{Conclusions}

In this paper we have shown that  
the KLM techniques can be extended to the case of polarization encoding.
This implementation has two main advantages. Firstly, it requires 
half the number of path modes when compared to the usual dual-rail
encoding. This can be very helpful in an actual implementation, since
a higher number of path modes makes it more difficult to
control and stabilize the interferometric setup required to implement the gates.
And secondly, when polarization encoding is used, the analogous to the basic KLM
scheme already has a photon loss detection mechanism. This is due to the fact that 
with this encoding we have one photon per path mode, instead of the 
one photon per two path modes of dual-rail encoding. Thus, when applying the non-deterministic
gates using photodetection, we know the total number of photons that are expected, 
and the information
about the operation of the gate is carried by the distribution of horizontally and
vertically polarized photons. 

The original KLM scheme can be modified to also include a photon loss
detection mechanism. This requires a  more complicated entangled ancilla that has
double the number of path modes. Polarization encoding not only reduces this number
by half, but also incorporates the loss mechanism in a natural way that makes it easier to 
understand how it works. It is just a consequence of the fact that the total
photon number is conserved by any linear-optical device. The construction of the 
entangled ancilla can proceed much in the same fashion as in the KLM case,
although the number of nondeterministic steps required roughly doubles. This 
is due to the fact that the polarization entangled ancilla have double the number
of photons. 

We have also studied numerically the effects of detector efficiency
on the basic KLM scheme and showed how important having a loss detection mechanism
is. Without it, even when the gate is assumed to be successful the probability
of errors can be as high as 30\%, and furthermore it increases with the number
of ancilla modes. Even though for perfect detectors the probability of success
of the gate increases with the number of ancillas, with imperfect detectors the probability
of the gate introducing no errors actually increases with the number of ancilla modes.
This shows that even for applications in which the success probability of the gate is
not required to be high (as in the construction of cluster states following Nielsen's
proposal), either loss detection or the use of high-efficiency detectors are crucial
for the success of the scheme.

Another interesting feature is that if the smallest entangled ancilla is used, which is 
a four-mode entangled state, the implementation of the CSIGN gate
becomes simpler. On one hand it does not require number-resolving
photodetectors, and on the other hand only a beam-splitter is required
to mix the input modes with the entangled ancila, instead of the
very complex interferometer required by the general Fourier transformation in
the KLM implementation.
Incidentally, it is worth noting that this mixing with a beam splitter
is very similar to the Type-II fusion operation introduced by  
Browne and Rudolph~\cite{browne2005a}, which can also be used to grow cluster
states. This operation also has the nice feature of its fidelity 
being independent of the detector efficiency (although this point 
was not mentioned by the authors.) This simplified implementation of the gate,
together with the fact that the output modes need not be physically post-selected
when the gate is successful, makes it very appealing for integration into an
optical chip.

It is important to note that even with the loss detection mechanism, 
the KLM scheme requires high-efficiency
detectors and high-fidelity single-photon sources in order to apply the nonlinear
sign gate that is the basis of the scheme. Here we have presented and alternative
approach that allows us to do away with high-efficiency detector and single-photon
source requirements, provided
we have access to a high-fidelity polarization-entangled Bell pair source.
This gives us another possible path to the implementation of LOQC
with low errors using currently availabe detectors.

\section{Acknowledgments}

Part of this work was carried out at the Jet Propulsion Laboratory, California
Institute of Technology, under a contract with the National Aeronautics and Space
Administration (NASA). FMS acknowledges support from the National Research Council
and NASA Code Y. JPD and HL acknowledge support from the Hearne Foundation, the
National Security Agency, the Advanced Research and Development Activity
and the Army Research Office.

\bibliographystyle{prsty}
\bibliography{Cav}

\end{document}